\documentclass{ws-procs961x669}            
\newcommand{\steady}[0]{\textit{16TI} }
\newcommand{\spikes}[0]{\textit{40sp\_down} }
\begin{document}
\title{Monte Carlo Simulations of Photospheric Emission in Gamma Ray Bursts}

\author{T. M. Parsotan$^*$ and D. Lazzati}

\address{Department of Physics, Oregon State University,\\
Corvallis, OR 97331, USA\\
$^*$E-mail: parsotat@umbc.edu}

\begin{abstract}
The study of Gamma Ray Bursts (GRBs) has the potential to improve our understanding of high energy astrophysical phenomena. In order to reliably use GRBs to this end, we first need to have a well-developed grasp of the mechanism that produces the radiation within GRB jets and how that relates to their structure. One model for the emission mechanism of GRBs invokes radiation produced deep in the jet which eventually escapes the jet at its photosphere. While this model has been able to explain a number of observed GRB characteristics, it is currently lacking in predictive power and in ability to fully reproduce GRB spectra. In order to address these shortcomings of the model, we have expanded the capabilities of the MCRaT code, a state of the art radiative transfer code that can now simulate optical to gamma ray radiation propagating in a hydrodynamically simulated GRB jet. Using the MCRaT code, we have constructed mock observed light curves, spectra, and polarization from optical to gamma ray energies for the simulated GRBs. Using these mock observables, we have compared our simulations of photospheric emission to observations and found much agreement between the two. Furthermore, the MCRaT calculations combined with the hydrodynamical simulations allow us to connect the mock observables to the structure of the simulated GRB jet in a way that was not previously possible. While there are a number of improvements that can be made to the analyses, the steps taken here begin to pave the way for us to fully understand the connection between the structure of a given GRB jet and the radiation that would be expected from it.
\end{abstract}


\bodymatter

\section{Introduction}\label{intro}
Gamma Ray Bursts (GRBs) typically produce high energy X-ray and gamma-ray radiation that is detected within the first few tens of seconds of the event. The origin of this so called prompt emission is still not well understood even after decades of investigation. The two most common models that are used to describe the prompt emission of GRBs either employ synchrotron emission from jets with magnetic fields \cite{ICMART_Zhang_2010, SSM_REES_MES, daigne1998_GRB_internal_shock, daigne2011marginally_fastcooling_synch} or emission from the photospheres of GRB jets \cite{REES_MES_dissipative_photosphere, Peer_fuzzy_photosphere, Peer_multicolor_bb, Peer_photospheric_non-thermal, Beloborodov_fuzzy_photosphere, lazzati_photopshere, MCRaT, parsotan_mcrat, parsotan_var, parsotan_polarization, ito2019photospheric}. 

Each model has its own advantanges and disadvantages allowing them to describe GRB observations at gamma-ray and X-ray energies in varying capacities. Models that rely on synchrotron emission are able to account for the non-thermal characteristics of GRB spectra \cite{daigne2011marginally_fastcooling_synch, Oganesyan2019_prompt_opt} but are unable to fully account for various GRB observational relationships such as the Amati \cite{Amati} and Yonetoku \cite{Yonetoku} relations \cite{ICMART_Zhang_2010}. The photospheric model on the other hand is able to reproduce the Amati and Yonetoku relations \cite{parsotan_mcrat, parsotan_var, lazzati_photopshere, lazzati_variable_photosphere, diego_lazzati_variable_grb, ito2019photospheric} but has issues with reproducing typical GRB spectral parameters \cite{parsotan_mcrat, parsotan_var}. In order to fully distinguish between these models additional data at different wavelengths is needed. 

There have been a number of optical prompt emission detections that can be used for the purpose of model comparison (see \cite{parsotan_optical} and references therein). Additionally, there has been one detection of optical prompt polarization \cite{troja2017_grb160625B} that can also be used in comparison with optical polarization predictions obtained from synchrotron and the photospheric model calculations. The synchrotron model has traditionally been favored by these optical prompt detections \cite{Racusin2008_grb080319B, shen_synch_optical_emission, Tang_2006_optical_lag, Fan_2009_neutron_proton_shells, Zou_2009_optical_different_region} however the photospheric model has not been fully explored in its ability to reproduce these optical prompt detections.

The ability to simulate radiation signatures produced under the photospheric model has recently been improved \cite{MCRaT, parsotan_mcrat, parsotan_var, Ito_3D_RHD, ito2019photospheric, parsotan_polarization, parsotan_optical, Parsotan_spectropolarimetry}. These analyses have combined the realistic treatment of radiation with hydrodynamically simulated time-dependent GRB jets. As a result, studies of how the GRB jet structure affects the detected radiation signatures are now possible. With the ability of these state-of-the-art radiative transfer calculations to also calculate the effects of cyclo-synchrotron emission and absorption \cite{parsotan_optical, Parsotan_spectropolarimetry}, these types of analyses are also now possible from optical to gamma-ray energies. In this proceeding, we highlight the results of conducting radiative transfer calculations of hydrodynamically simulated GRB jets and the construction of mock observations from these calculations. We show how the mock observables can be related to the structure of the simulated GRB jets and how these analysis can help us interpret GRB observations under the photospheric model. Finally, we conclude with a summary and highlight ways that these types of global radiative transfer calculations can push the field forward.

\section{Methods}
\subsection{The MCRaT Code}

\begin{figure}[t!]
\begin{center}
\includegraphics[width=\textwidth]{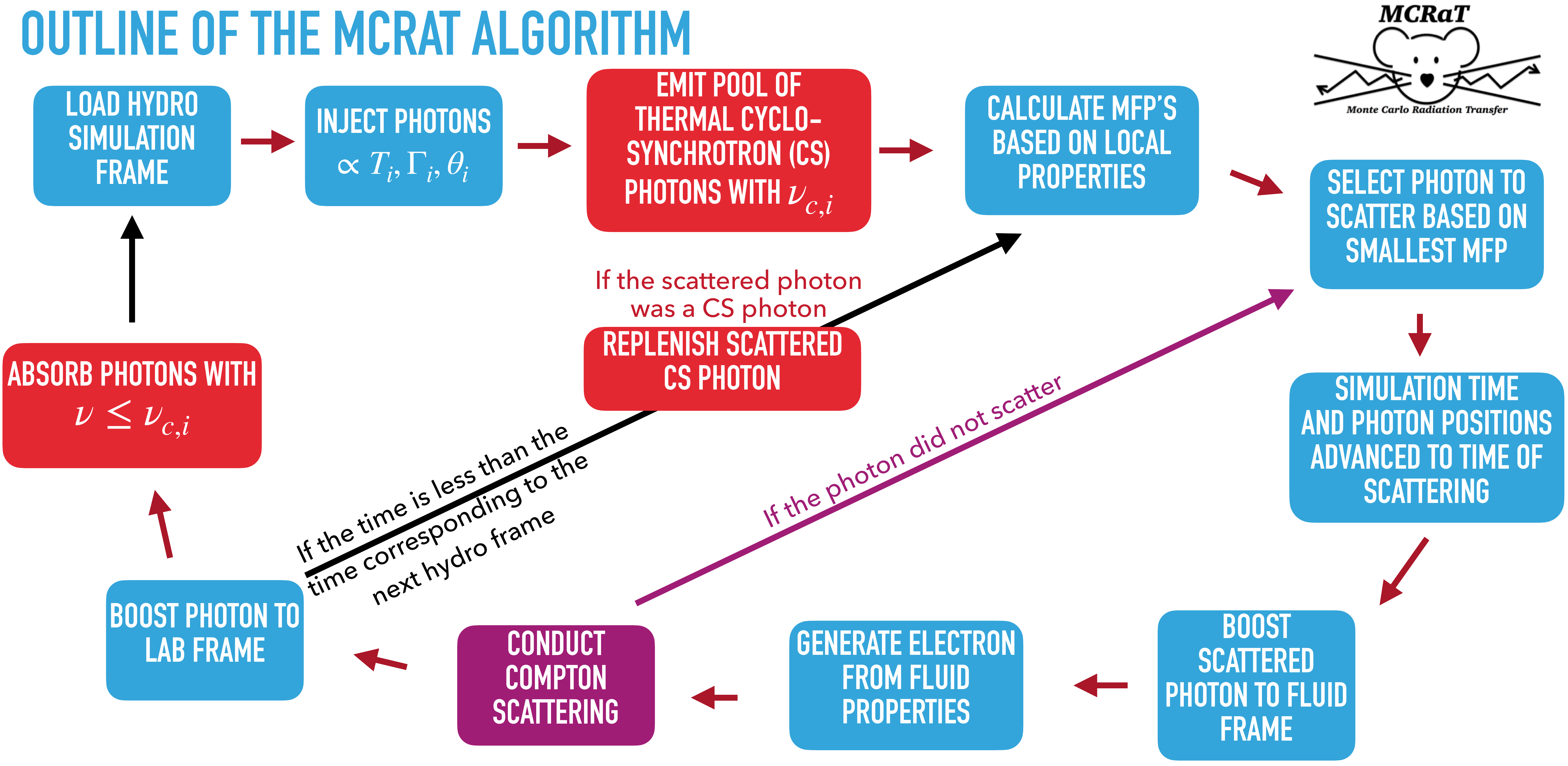}
\end{center}
\caption{The outline of the MCRaT code, see text for in depth explanation.}
\label{mcrat_algorithm}
\end{figure}

We have used the open-source {\bf{M}}onte {\bf{C}}arlo {\bf{Ra}}diation {\bf{T}}ransfer (MCRaT) code\cite{parsotan_mcrat_software_2021_4924630} to analyze the evolution of radiation trapped within hydrodynamical simulations of GRB jets. The MCRaT code considers photons in the astrophysical outflow that interact via Compton scattering including the effects of the full Klein-Nishina cross section with polarization. The code also considers photons that have been absorbed and emitted in the outflow due to cyclo-synchrotron (CS) emission and absorption. We summarize the algorithm here, however detailed descriptions of the code are outlined in \cite{parsotan_optical} and \cite{Parsotan_spectropolarimetry}. As is outlined in Figure \ref{mcrat_algorithm}, the code loads a SRHD simulation frame and injects photons into the simulated GRB jet where the optical depth $\tau > 100$. The code will calculate the mean free paths of each photon in the simulation in order to determine which will scatter. After this calculation, MCRaT advances the simulation time to correspond to when the scattering will occur and propagates each photon by a distance corresponding to the amount of time passing in the simulation. The code then conducts the scattering considering the full Klein-Nishina cross section with polarization, which means that the photons have some probability of not scattering in the electron rest frame. If the chosen photon ends up not being scattered, MCRaT chooses the next photon that would have scattered and advances the simulation time and the location of each photon. If the photon is scattered, then the new four momentum and polarization of the photon is calculated. The code continues to scatter photons until the time in MCRaT corresponds to the time in the next SRHD simulation frame. If CS emission and absorption is also taken into account in the simulation then the steps in the red boxes are taken which numerically simulate CS emission and absorption processes in MCRaT.

\subsection{Mock Observations}
With a given MCRaT simulation completed, we use the open-source ProcessMCRaT code\cite{parsotan_processmcrat_software_2021_4918108} to produce mock observed light curves, spectra, and polarization measurements that can be compared to actual GRB observations on a more equal basis. This code allows photons from an MCRaT simulation to be binned in time and energy in order to produce the mock observables, similar to how actual photons measured in a given GRB event are binned. Detailed descriptions of how we construct the mock observable quantities for observers at various viewing angles with respect the jet axis, $\theta_\mathrm{v}$, can be found in \cite{parsotan_mcrat, parsotan_var, parsotan_polarization, parsotan_optical}. We use the same gamma-ray and optical energy ranges as in \cite{Parsotan_spectropolarimetry}. The gamma-ray mock observables are calculated from photons with energies between 20-800 keV which corresponds to the polarimetry energy range of POLAR-2 \cite{hulsman2020polar2}. The optical mock observables are calculated from photons with energies between 1597-7820 $\AA$ ($\sim 1.5-7.7$ eV), which aligns with the Swift UVOT White bandpass \cite{poole2008_swiftphotometric,rodrigo2020svo}.

The way that we connect the mock observables to the structure of the simulated GRB jets is through equal arrival time surfaces (EATS) \cite{Zhang_E_p_evolution, Peer_multicolor_bb, beloborodov2011radiative} of the photons in the GRB jet. For a given time bin, in the mock observed light curve, we can calculate surfaces that would be emitting photons towards a given observer's line of sight. These surfaces are calculated following the method of \cite{parsotan_polarization}, who describe the procedure for constructing these EATS.

\subsection{The Simulation Set}
The MCRaT code was used to analyze a set of two special relativistic hydrodynamical simulations (SRHD) of GRB jets. These SRHD simulations were conducted using the FLASH hydrodynamics code \cite{fryxell2000flash} and their properties are outlined in \cite{parsotan_polarization, parsotan_optical, Parsotan_spectropolarimetry}. These two SRHD simulations are referred to as the \steady and \spikes simulations and they represent GRB jets that have a  time-constant and time-variable injection of energy, respectively. 

\section{Results}
\begin{figure*}
\includegraphics[width=\textwidth]{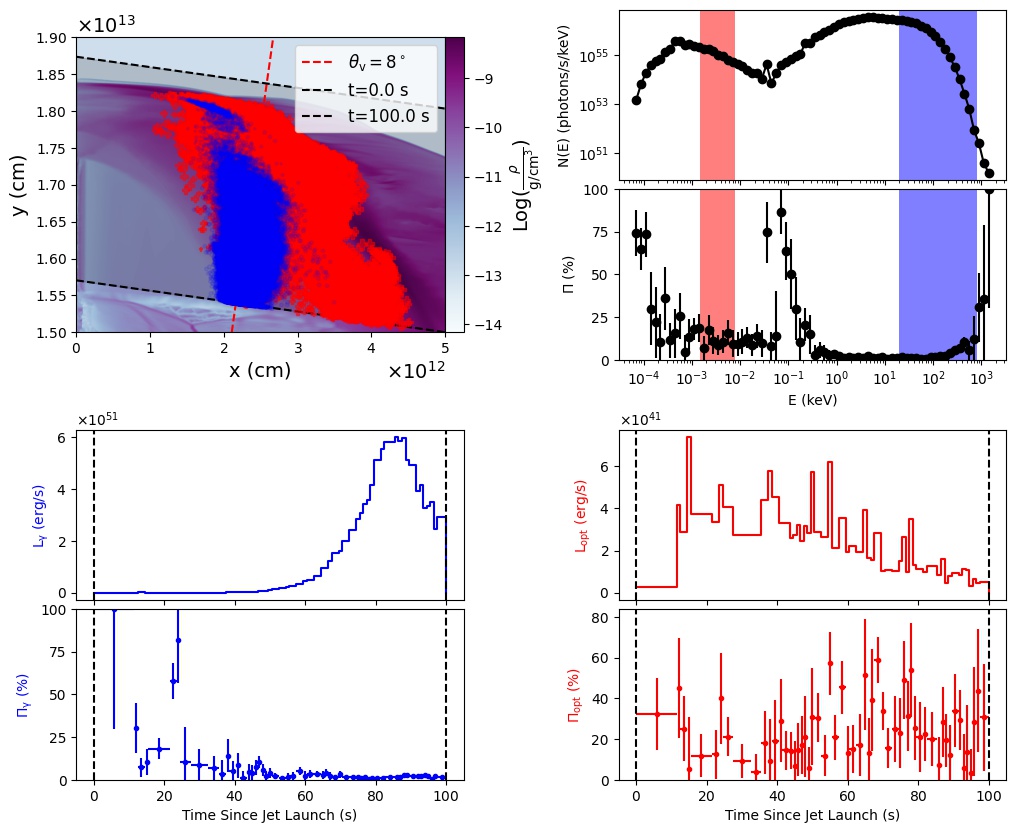}
\caption{The relation between the mock observed quantities and the jet structure of the \steady simulation for $\theta_\mathrm{v}=8^\circ$. See text for full description. 
}
\label{16ti_ani}
\end{figure*}


\begin{figure}[]
 \centering
 \includegraphics[width=\textwidth]{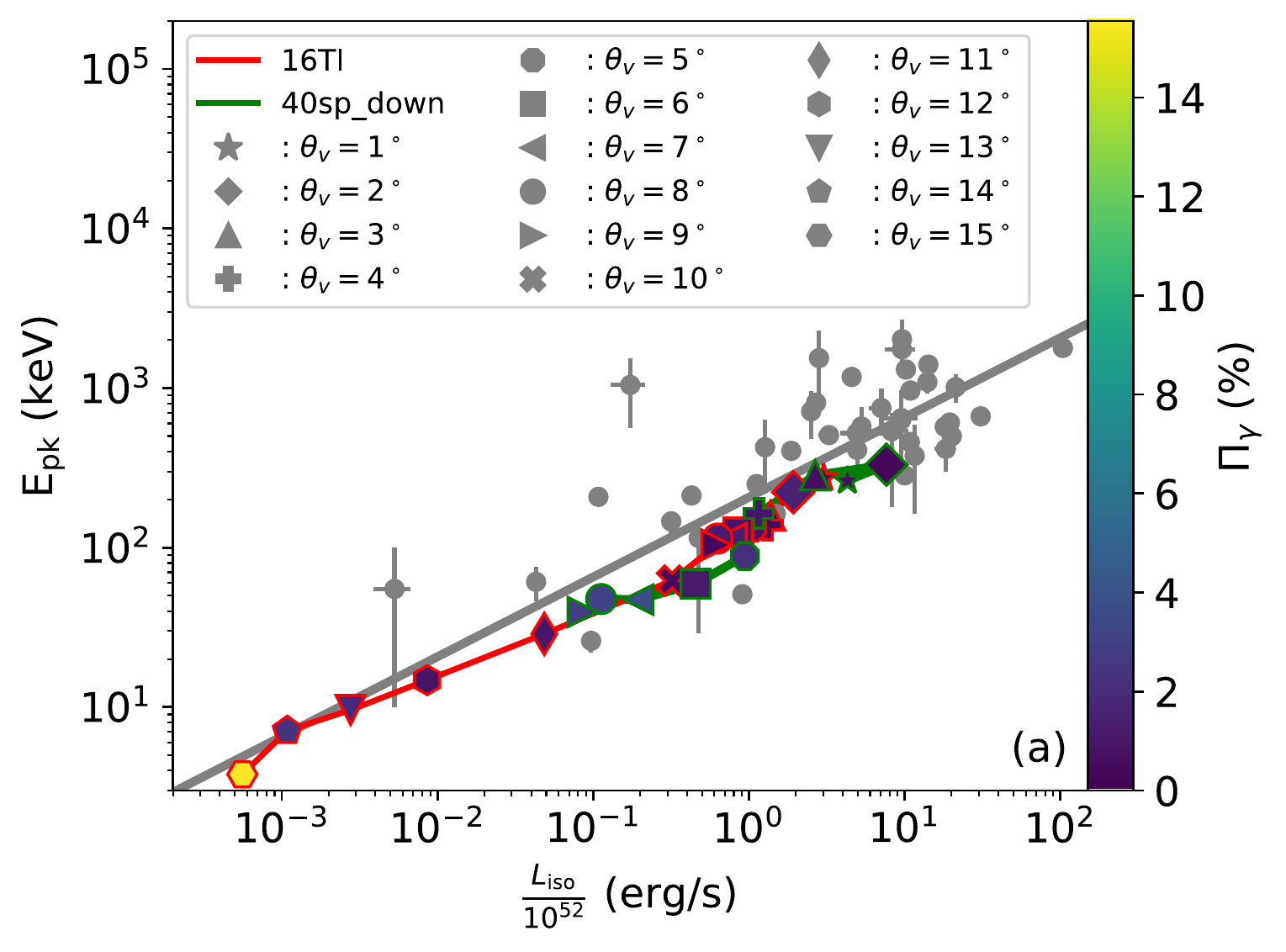}
 \includegraphics[width=\textwidth]{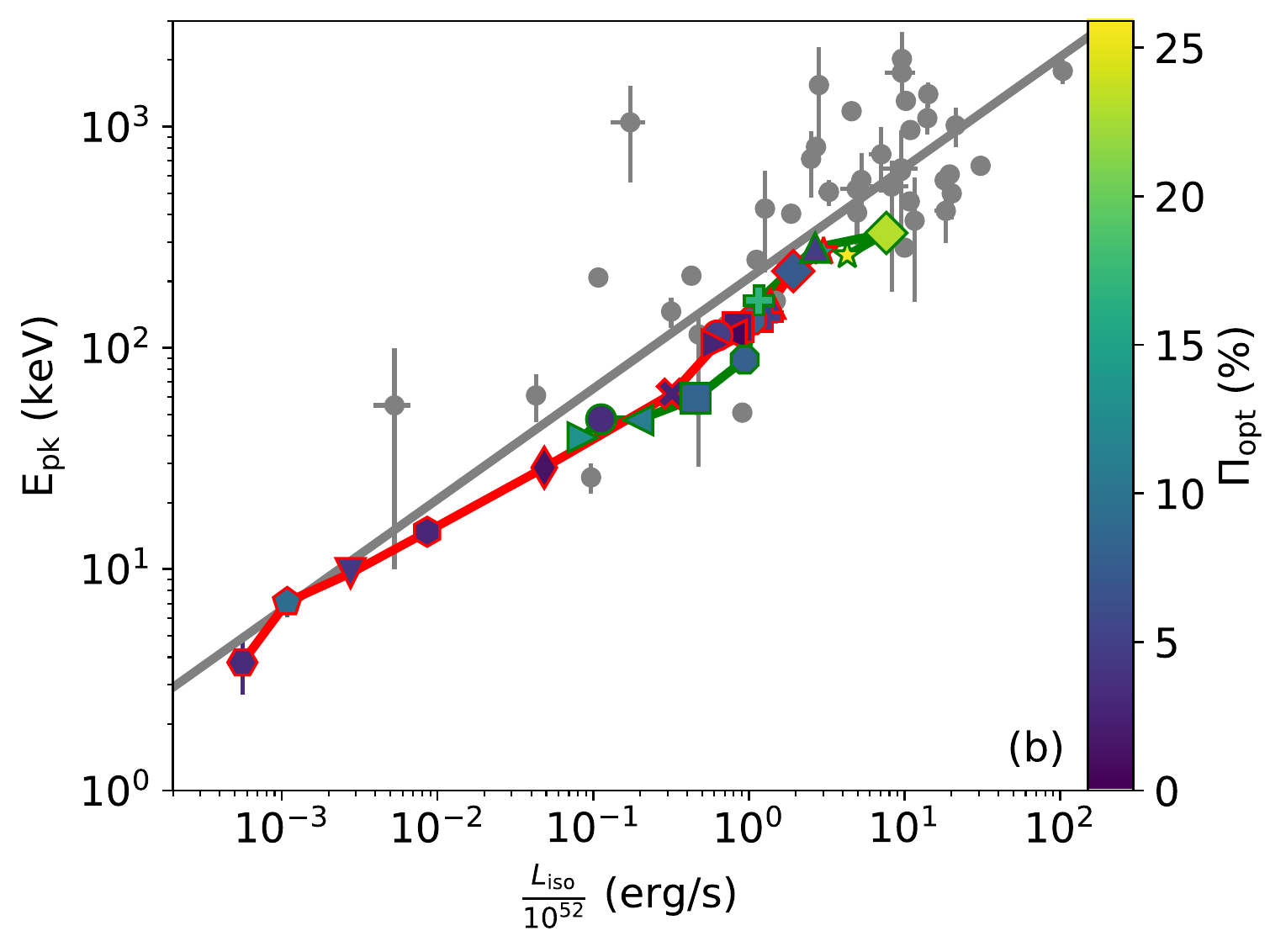}
 \caption{The \steady and \spikes simulations plotted alongside the Yonetoku relationship \cite{Yonetoku} in red and green lines and marker outlines, respectively. In Figure (a), the fill color of the marker denotes the mock observed time-integrated $\Pi_\gamma$ while in Figure (b) the fill color shows the time-integrated MCRaT $\Pi_\mathrm{opt}$. The different marker shapes for the MCRaT simulations show the placement of the simulations on the Yonetoku relationship as determined by observers at various $\theta_\mathrm{v}$. In each panel the observational relationship is shown as the grey solid line and observed data from \cite{data_set} are plotted as grey markers. Figure adapted from \cite{Parsotan_spectropolarimetry}.} 
 \label{yonetoku_plot}
\end{figure}

Our MCRaT simulations have shown that optical and gamma ray photons probe different regions of GRB outflows under the photospheric model, as is shown in Figure \ref{16ti_ani}. Here, we show all the photons that have been detected by an observer located at $\theta_\mathrm{v}=8^\circ$ in optical and gamma-ray energies. The top left panel shows the location of each photon in relation to the density structure of the SRHD simulated GRB jet. The density of the jet is shown as a pseudocolor plot where darker colors show denser regions and the dashed red line denoted the line of sight of the observer at $\theta_\mathrm{v}=8^\circ$. The optical photons are shown in red while gamma-ray photons are depicted in blue. Furthermore, the photon markers are translucent allowing us to identify regions of the jet where the photons are densely located (due to the concentration of blue or red). Additionally, the markers are different sizes in order to show the weight of each photon in the calculation of the various mock observable quantities. The top right panel shows the time integrated spectrum of these photons with highlighted red and blue regions that highlight the energy ranges which were used to calculate the optical and gamma-ray light curves ($L_\gamma$ and $L_\mathrm{opt}$) and polarizations ($\Pi_\gamma$ and $\Pi_\mathrm{opt}$). These time-resolved mock observables are shown in the bottom 4 panels. These analysis allow us to connect the mock observables of light curves, spectra, and polarizations with the locations of where the photons are within the jet and what the jet structure is at those locations. As is shown in Figure \ref{16ti_ani}, the optical photons are distrubuted widely about the observer's line of sight with their locations being focused at the dense jet-cocoon interface (JCI;\cite{gottlieb2021structure}). The gamma-ray photons instead are mostly concentrated along the observer's line of sight with the exception of some gamma-ray photons being located near the core of the jet. Thus, optical prompt emission probes the dense JCI while the gamma-ray emission probes the portion of the jet that is directly along the observer's line of sight. 

The results of our MCRaT simulations can also be used to assess how well the simulations reproduce various observational relationships, such as the Yonetoku relation \cite{Yonetoku}, and make additional predictions that can be tested against data. Figure \ref{yonetoku_plot} shows the comparison between the \steady and \spikes simulations, in red and green marker edge colors and lines respectively, and the Yonetoku relation. Various observed GRBs from \cite{data_set} are also shown as grey circles. We also show the time-integrated $\Pi_\gamma$ and $\Pi_\mathrm{opt}$ in Figure \ref{yonetoku_plot}(a) and (b) respectively, through the fill-color of each marker. In each panel, we see that both the \spikes and \steady simulations are able to recover the normalization and slope of the Yonetoku relation for a variety of $\theta_\mathrm{v}$. With this success of the photospheric model, we can also now use the simulated polarization mock observations to make predictions. We find that the time-integrated $\Pi_\gamma$ should increase as $\theta_\mathrm{v}$ increases which has been found in other studies \cite{ito_polarization, lundman2014polarization}. Thus, bursts that are bright and energetic (in the top right of the Yonetoku relation) should have low time-integrated $\Pi_\gamma$. On the other hand, the time-integrated $\Pi_\mathrm{opt}$ should be larger for $\theta_\mathrm{v}$ that are close to the jet axis. As a result, we would expect that bursts located in the top right of the Yonetoku relation would have large time-integrated $\Pi_\mathrm{opt}$.

\section{Conclusion}
Using the MCRaT radiative transfer code, we have conducted post-processing analysis of the radiation expected from time-dependent special relativistic hydrodynamic simulations of GRB jets. These global radiative transfer simulations allow for the connection between observed radiation signatures and the structure of the GRB jet, which was not possible prior to these types of analysis. These analyses are commonplace in other areas of astrophysics such as galaxy formation and have proved fruitful in connecting radiation processes to the structure of galaxies and in testing observational data analysis techniques (see e.g. \cite{parsotan2021realistic} and \cite{cochrane2019predictions}). While these types of analysis are beginning to emerge in the study of GRBs (see references within), we need widespread adoption of these simulations within the community for any SRHD simulations that have been conducted of GRB jets. This would allow the community to build a library of GRB jet simulations and their associated radiation signatures that can be used in comparison to GRB observations, greatly enhancing our knowledge of GRB jets and the prompt emission.

\bibliographystyle{ws-procs961x669}
\bibliography{references}

\end{document}